# High-order Radio Frequency Differentiation via Photonic Signal Processing with an Integrated Micro-resonator Kerr Optical Frequency Comb Source

Xingyuan Xu,[1] Jiayang Wu,[1] Mehrdad Shoeiby,[2] Sai T. Chu,[3] Brent E. Little,[4] Roberto Morandotti,[5] Arnan Mitchell,[2] and David J. Moss[1,*]

[1]*Centre for Micro-Photonics, Swinburne University of Technology, Hawthorn, VIC 3122 Australia*
[2]*School of Engineering, RMIT University, Melbourne, VIC 3000, Australia*
[3]*Department of Physics and Material Science, City University of Hong Kong, Hong Kong, China.*
[4]*Xi'an Institute of Optics and Precision Mechanics Precision Mechanics of CAS, Xi'an, China.*
[5]*INRS – Énergie, Matériaux et Télécommunications, Varennes, Québec, J3X 1S2, Canada.*
[*dmoss@swin.edu.au](mailto:dmoss@swin.edu.au)

*Abstract* — We demonstrate the use of integrated micro-resonator based optical frequency comb sources as the basis for transversal filtering functions for microwave and radio frequency photonic filtering and advanced functions.

*Keywords—frequency comb, microwave, micro-resonator*

## I. Introduction

Photonic integrated circuits that exploit nonlinear optics for all-optical signal processing have been demonstrated, particularly in silicon, including all-optical logic [1], demultiplexing from 160Gb/s [2] to over 1Tb/s [3], to optical performance monitoring using slow light at speeds of up to 640Gb/s [4-5], all-optical regeneration [6], and many others [7-16], most based on third order nonlinearities including the Kerr nonlinearity as well as optical third harmonic generation (THG) [17-18]. The efficiency of all-optical devices depends on the waveguide nonlinear parameter, $\gamma = \omega\, n_2 / c\, A_{eff}$. Although silicon can achieve extremely high values of $\gamma$, it suffers from high nonlinear losses due to two-photon absorption (TPA) and the resulting free carriers. Even if the free carriers are eliminated by p-i-n junctions, silicon's poor intrinsic nonlinear figure of merit (FOM = $n_2 / (\beta\, \lambda)$, where $\beta$ is the TPA) of around 0.3 in the telecom band is far too low to achieve high performance. While TPA can be turned to advantage for all-optical functions [19-21], for the most part silicon's low FOM in the telecom band is a limitation.

In 2008-2010, new CMOS compatible platforms for nonlinear optics were introduced, including Hydex [22-30] and silicon nitride [31]. These platforms exhibit negligible nonlinear absorption in the telecom band, and have revolutionized micro-resonator optical frequency combs. The first integrated CMOS compatible integrated optical parametric oscillators were reported in 2010 [23, 31], showing that Kerr frequency comb sources could be realised in chip form by using ring resonators with relatively modest Q-factors. Following this, a stable modelocked laser with pulse repetition rates from 200GHz to 800GHz was reported [24]. The success of this platform arises from its very low linear loss, its moderate nonlinearity ($\gamma \cong 233\,\text{W}^{-1}\text{km}^{-1}$) and negligible nonlinear loss (TPA) [30]. This is critical since, even some of the best nonlinear platforms have only achieved a moderately high FOM [32-34] which techniques such as high Q cavities and slow light are not able to overcome [35-37].

Here, we review recent results for radio-frequency (RF) and microwave photonics applications of micro-combs. Optical frequency combs, and microcombs in particular, are ideal sources with which to base microwave and RF transversal filters on [38-44]. Many RF applications, including radar mapping, measurement, imaging as well as the realization of advanced modulation formats for digital communications, require the generation, analysis and processing of analogue RF signals where both the amplitude and phase of the signals are important. In-phase and quadrature-phase components of a signal can be obtained via a photonic based Hilbert transform [39].

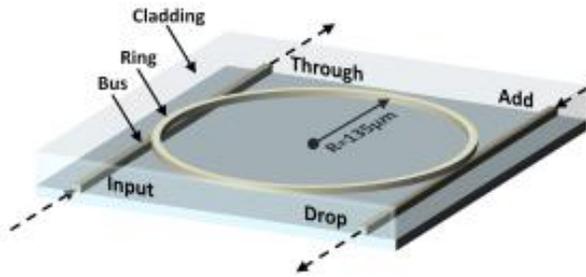

Fig. 1. Integrated ring resonator.

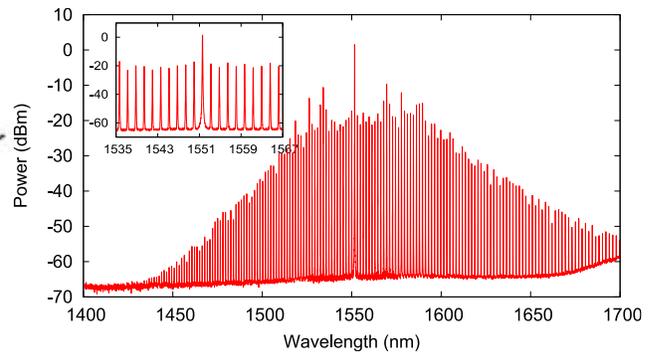

Fig. 2. Optical frequency comb generated by microresonator.

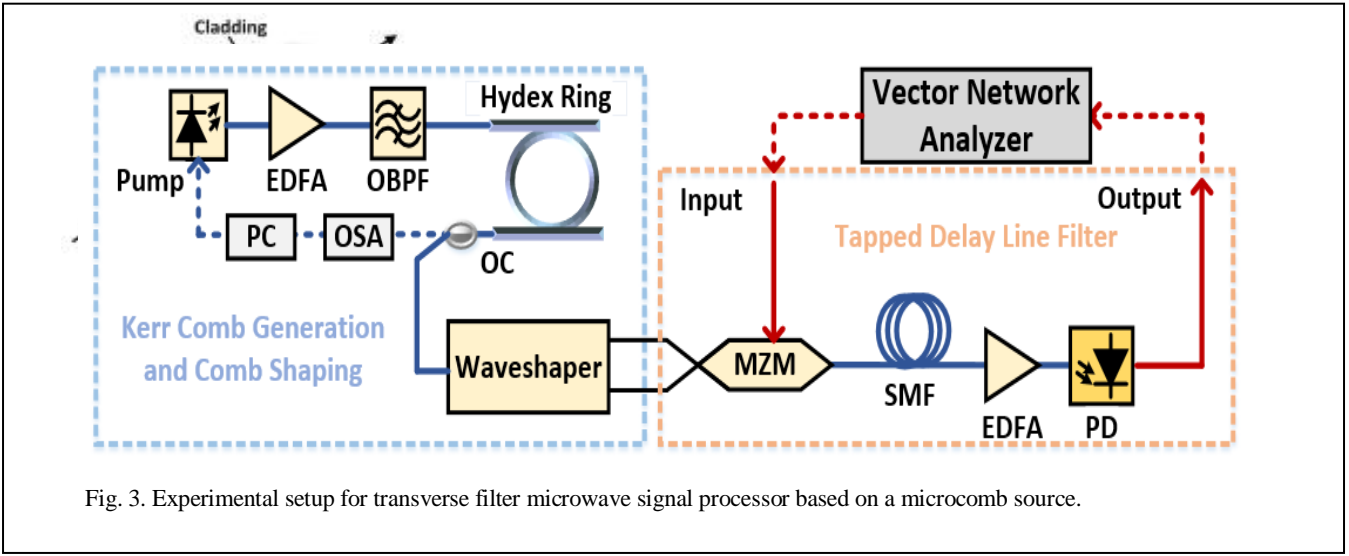

Fig. 3. Experimental setup for transverse filter microwave signal processor based on a microcomb source.

II. EXPERIMENT

Figure 1 shows a schematic of our ring resonator, with a Q factor of 1.2 million and FSR of 200GHz [23]. The resulting comb spectrum is shown in Figure 2 and in more detail in Figure 4, and had a spacing corresponding to the FSR of the microring. It was a "Type II" comb [45, 46] with limited coherence between the comb lines. For the photonic microwave signal processing functions studied here, however, this was adequate.

Figure 3 illustrates the experiment setup consisting of frequency comb generation, comb shaping, and photonic RF filtering. A continuous-wave (CW) tunable laser (Agilent 8160A), amplified by a high power Erbium-doped fiber amplifier (EDFA), was used as the pump source for the high-Q microring resonator. The equally spaced comb lines produced by the ring resonator were then shaped according to the required tap coefficients using a reconfigurable filter (a Finisar S400 waveshaper), having a much smaller resolution than the comb line spacing. The waveshaper also split the comb into two paths, which were connected to the photonic RF filter section. Comb lines corresponding to positive tap coefficients were routed to one of the output ports of the waveshaper, while comb lines corresponding to negative tap coefficients were sent to the second port.

The two outputs of the waveshaper were connected to two inputs of a 2x2 balanced Mach-Zehnder modulator (MZM), biased at quadrature, where the comb lines were modulated by the RF signal. One group of comb lines was modulated on the positive slope of the MZM transfer function while the second group was modulated on the negative slope. This allowed both negative and positive tap coefficients to be realized with a single MZM. The output of the MZM was then passed through 2.122 km of single-mode fiber (SMF) that acted as a dispersive element to delay the different filter taps. The dispersed signal was then amplified by a second fiber amplifier to compensate for loss, and filtered in order to separate the comb from the signal at the pump wavelength in order to produce the system 0° phase reference.

The second path was used as the 90° phase signal. The signal path was time-shifted with a variable optical length (VOL) so that the reference could be positioned, in time, exactly at the middle of the filter taps. The optical signals were finally detected by photodiodes to produce the output RF signals. The system RF frequency response was then measured with an RF vector network analyzer (VNA).

The resulting Type II Kerr optical comb [46] as shown in Fig. 4(a) is over 200-nm wide, and flat over ~32 nm. Since the generated comb only served as a multi-wavelength source for the subsequent transversal filter in which the optical power from different taps was detected incoherently by the photo-detector, the coherence of the comb was not crucial and the proposed differentiator was able to work under relatively incoherent conditions. In the experiment, the numbers of taps used for fractional-, first-, second-, and third-order differentiation demonstrations were 7, 8, 6, and 6, respectively. The choice of these numbers was made mainly by considering the power dynamic range of the comb lines, i.e., the difference between the maximum power of the generated comb lines and the power associated with the noise floor. The power dynamic range was determined by the EDFA before waveshaping, and in our case, it was ~30 dB.

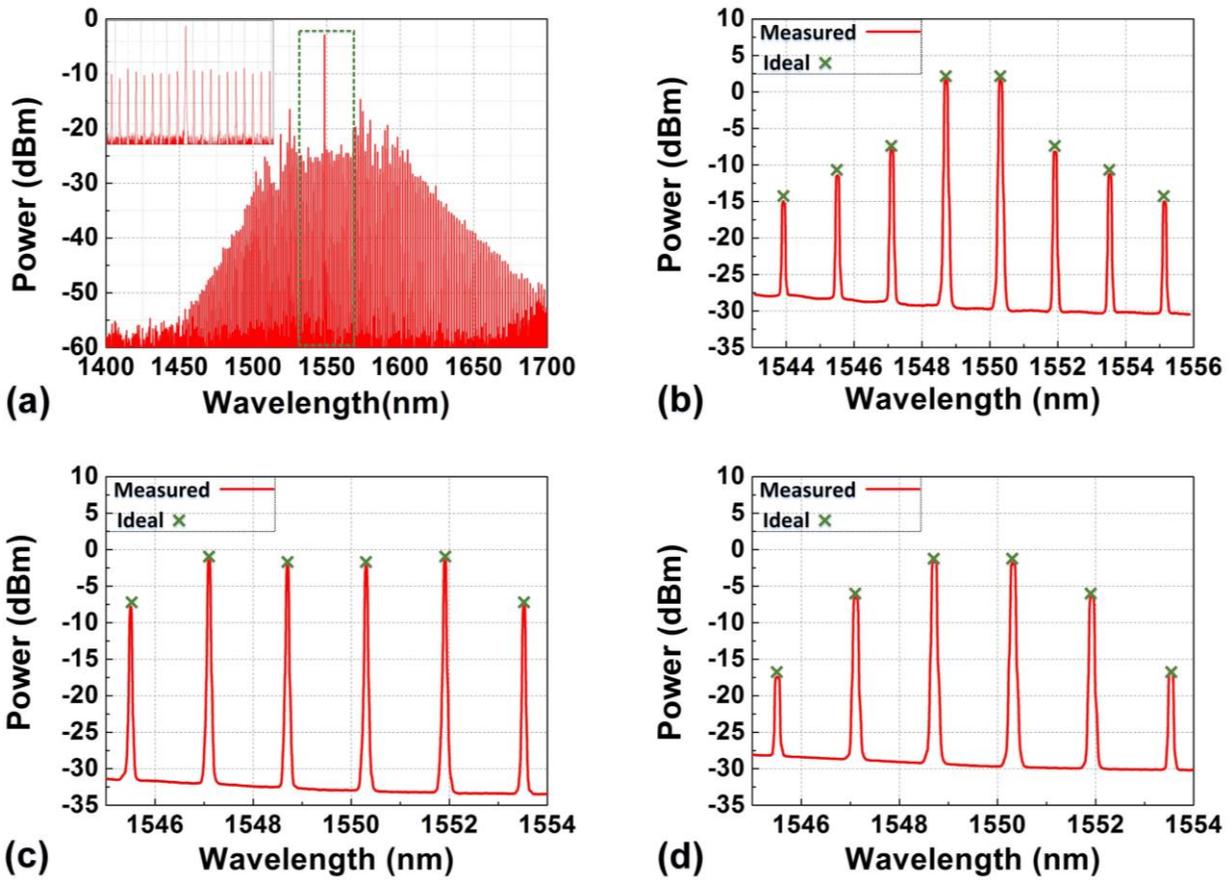

Fig. 4. (a) Optical spectrum of the generated Kerr comb in a 300-nm wavelength range. Inset shows a zoom-in spectrum with a span of ~32 nm. (b)–(d) Measured optical spectra (red solid) of the shaped optical combs and ideal tap weights (green crossing) for the first-, second-, and third-order intensity differentiators.

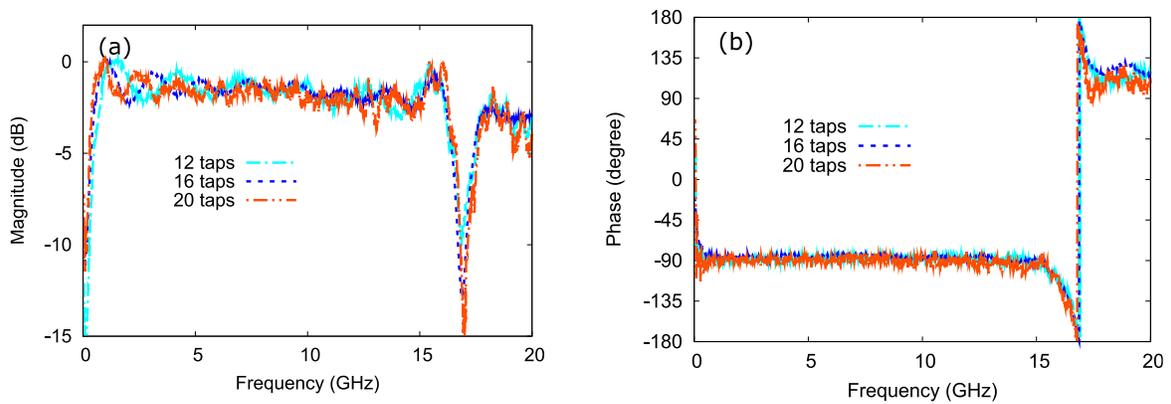

Fig. 5. Amplitude and phase response of Hilbert Transformer

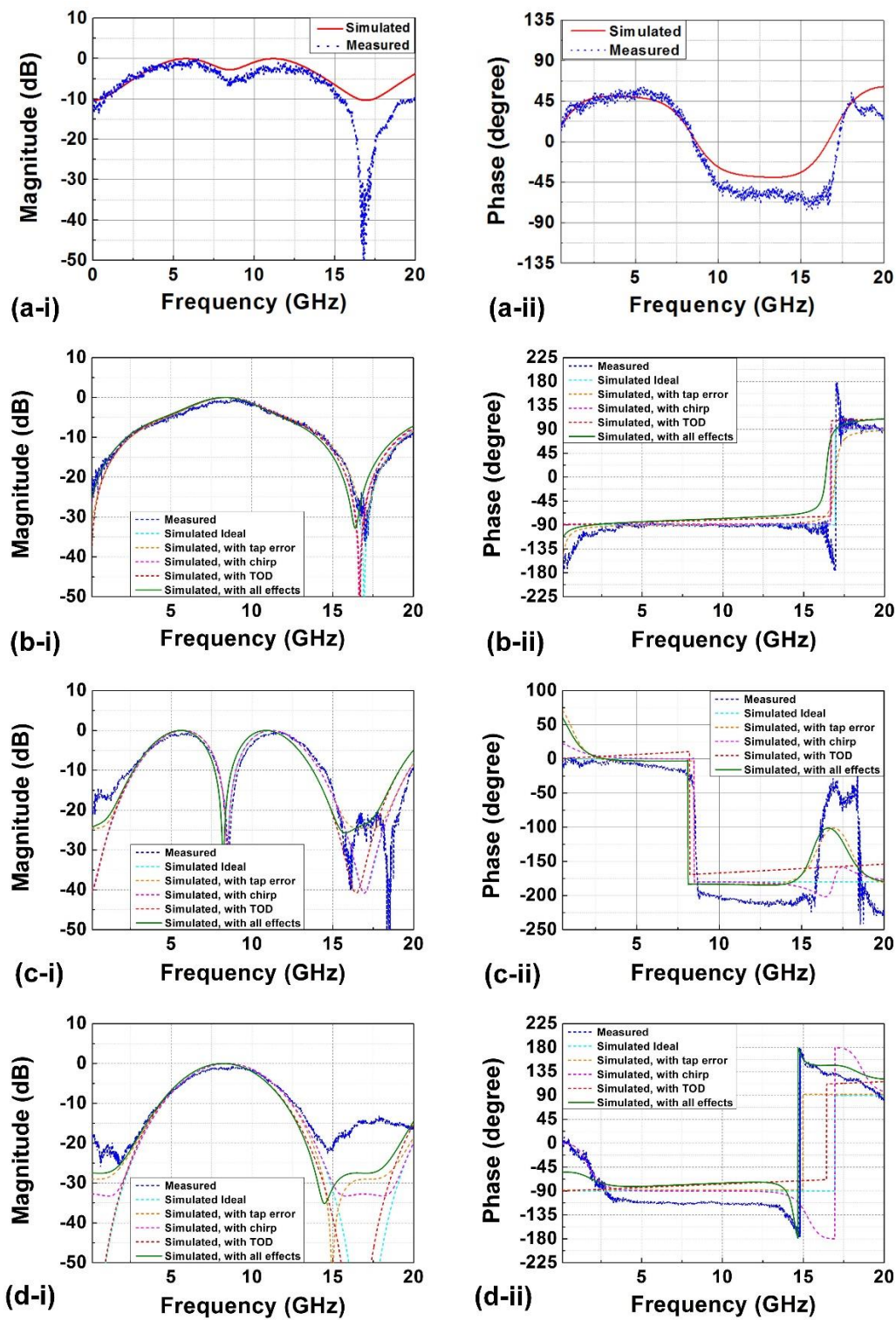

Fig. 6. Measured and simulated RF amplitude and phase responses of (a-i)–(a-ii) the fractional-order, (b-i)–(b-ii) the first-order, (c-i)–(c-ii) second-order, and (d-i)–(d-ii) third-order intensity differentiators. The simulated amplitude and phase responses after incorporating the tap error, chirp, and the third-order dispersion (TOD) are also shown accordingly.

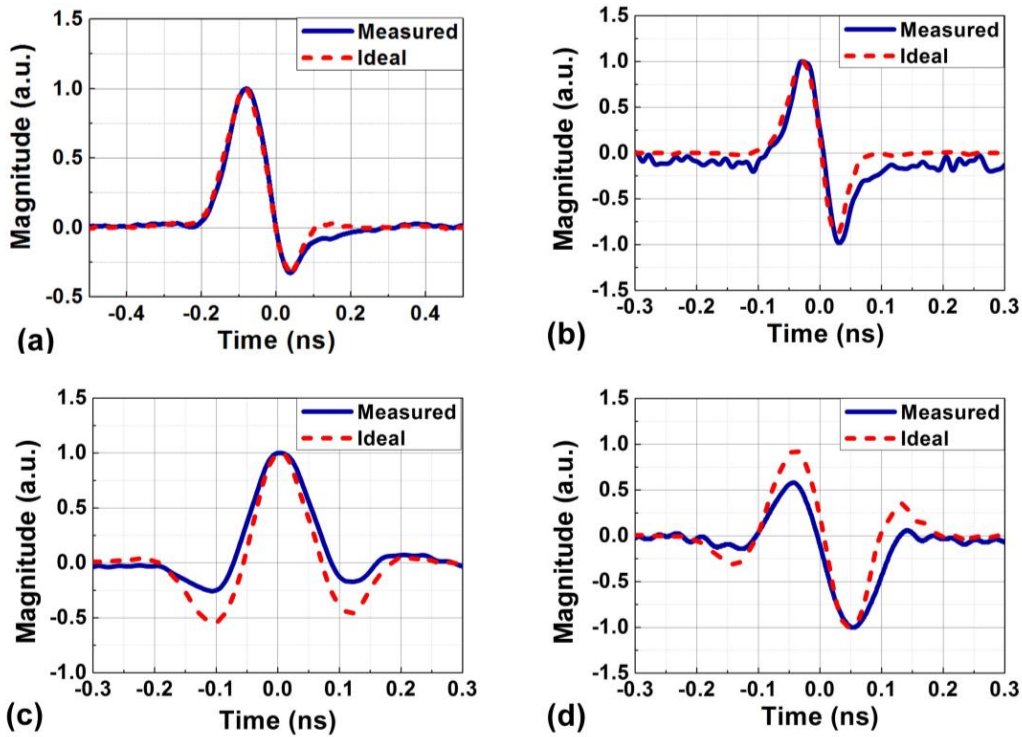

Fig. 7. Theoretical (red dashed) and experimental (blue solid) responses of the (a) fractional- (b) first-, (b) second-, and (c) third-order intensity differentiators.

# Table 1.

Tap coefficients for the fractional-, first-, second-, and third-order differentiations.

| Order of differentiation | Number of taps | Tap coefficients |
| --- | --- | --- |
| Fractional-order | 7 | [0.0578, -0.1567, 0.3834, 1, -0.8288, 0.0985, -0.0892] |
| First-order | 8 | [−0.0226, 0.0523, −0.1152, 1, −1, 0.1152, −0.052, 0.0226] |
| Second-order | 6 | [0.0241, −0.1107, 0.0881, 0.0881, −0.1107, 0.0241] |

## III. RESULTS AND DISCUSSION

Figure 5 shows the measured RF amplitude frequency response of a photonic Hilbert transform filter for 12, 16 and 20 taps, respectively, which all exhibit expected behavior. All three filters had the same null frequency at 16.9 GHz, corresponding to the tap spacing of $\Delta t = 1/f_c = 59$ ps. This spacing matches the difference in delay between the comb lines, equal to the ring $FSR = 1.6$ nm, produced by propagation through a 2.122 long SMF fiber with a dispersion parameter $D = 17.4$ ps/(nm km). The null frequency could be controlled by using a different fiber length to adjust the tap spacing.

All filters show < 3 dB amplitude ripple. Increasing the number of filter taps increases the filter bandwidth. With a 20 tap filter, the Hilbert transformer exhibited a 3 dB bandwidth extending from 16.4 GHz down to 0.3 GHz, corresponding to more than 5 octaves. It is possible to increase this bandwidth further by using more comb lines in the filter. In our experiment, only a small portion of the generated comb spectrum was actually used to realize the filter taps. The number of filter taps that could be achieved was actually limited by the bandwidth of the waveshaper and the gain bandwidth of the optical amplifier. If desired, it would also be possible to reduce the amplitude ripple within the pass-band by apodizing the tap coefficients from the ideal hyperbolic function [39].

Figure 5 shows the measured phase response of filters with different numbers of taps, showing very similar responses. Each shows a relatively constant phase of near -90º within the pass-band. There are some deviations from the ideal -90º phase at frequencies close to zero and particularly for the null frequency $f_c$ = 16.9 GHz.

In order to demonstrate differentiation functions [38], eight comb lines were selected and shaped by the waveshaper with tap coefficients shown in Table I [-0.0226, 0.0523, -0.1152, 1, -1, 0.1152, -0.052, 0.0226]. As before, the shaped comb lines were then divided into two parts according to the tap coefficients and fed into the 2×2 balanced MZM biased at quadrature, and then propagated through 2.122-km of single mode fibre with a dispersion of 17.4 ps/(nm km), corresponding to a time delay of ~59 ps between adjacent taps and a yielding an effective FSR of ~16.9 GHz.

After the weighted and delayed taps were combined upon detection, the RF responses for different differentiation orders were characterized by a vector network analyser (VNA, Anritsu 37369A). Figures. 6(a-i), (b-i), (c-i) and (d-i) show the measured and simulated amplitude responses of the first-, second-, and third-order intensity differentiators, respectively. The corresponding phase responses are depicted in Figs. 6(a-ii), (b-ii), (c-ii) and (d-ii). It can be seen that all the three differentiators feature the responses expected from ideal differentiations. The FSR of the RF response spectra is ~16.9 GHz, which is consistent with the time delay between adjacent taps, as calculated before measurement. Note that by adjusting the FSR of the proposed transversal filter through the dispersive fibre or by programming the tap coefficients, a variable operation bandwidth for the intensity differentiator can be achieved, which is advantageous for meeting diverse requirements in operation bandwidth.

We also performed system demonstrations of real-time signal differentiations for Gaussian input pulses with a full-width at half maximum (FWHM) of ~0.12 ns, generated by an arbitrary waveform generator (AWG, KEYSIGHT M9505A). The waveform of the output signals after differentiation are shown in Figs. 7(a)–(d) (blue solid curves). They were recorded by means of a high-speed real-time oscilloscope (KEYSIGHT DSOZ504A Infiniium). For comparison, we also depict the ideal differentiation results, as shown in Figs. 7(a)–(d) (red dashed curves). The practical Gaussian pulse was used as the input RF signal for the simulation. One can see that the measured curves closely match their theoretical counterparts, indicating good agreement between experimental results and theory. Unlike the field differentiators, temporal derivatives of intensity profiles can be observed, indicating that intensity differentiation was successfully achieved. For the first-, second-, and third-order differentiators, the calculated RMSE between the measured and the theoretical curves are ~4.15%, ~6.38%, and ~7.24%, respectively.

## IV. CONCLUSIONS

In summary, we show that an integrated optical comb source can be effectively used to provide numerous, high quality optical taps for a microwave photonic transversal filter, thus allowing us to demonstrate very wide bandwidth RF filtering and signal processing functions with a 3 dB bandwidth of over 5 octaves from as low as 0.3 GHz to 16.9 GHz. It is extremely difficult to match this performance using electronic or other photonic techniques. In our work, only a relatively small number of the available comb lines from the integrated comb source were utilized to realize the filter taps. This was limited by the finite bandwidth of the configurable filter used to shape the comb spectrum as well as the optical fiber amplifiers. Reducing loss in order to potentially eliminate the amplifier, as well as using a wider bandwidth configurable filter, will both allow more comb lines to be used, resulting in an even broader RF bandwidth. Further improvements to the filter ripple and response near the band edges can be achieved through apodization and compensation of imperfections in the modulation and transmission system. Since the integrated comb source can generate many more comb lines than the number of filter taps, it is possible to realize multiple parallel filters using only a single comb source, further reducing the device complexity.

We demonstrate a range of functions for microwave and RF signal processing based on photonic transverse filters using an integrated microcomb multiple wavelength source. We demonstrate a wideband Hilbert transformer as well as differentiation functions using a transversal filtering scheme. The wide spectral width and large frequency spacing of the integrated comb source allows a large number of high quality filter taps without any increase in the system complexity, thus achieving a much wider RF bandwidth than what can be typically obtained with standard microwave circuits. We achieve more than a 5-octave bandwidth while maintaining a nearly frequency independent quadrature phase over the pass-band. Our approach also has the potential to achieve full integration on a chip, which could have a significant impact on signal processing systems for many applications including radar detection, imaging and communications.

ACKNOWLEDGMENTS

This work was supported by the Australian Research Council Discovery Projects scheme.

`